\documentclass[review]{elsarticle}
\usepackage[utf8]{inputenc}
\usepackage{url}
\usepackage{amsmath,amssymb,amsfonts,dsfont, graphicx}
\usepackage{stmaryrd}
\usepackage{multirow}
\usepackage[table,xcdraw]{xcolor}
\usepackage[legalpaper, portrait, margin=2cm]{geometry}

\graphicspath{{figures/}, {./}}

\def \R {\mathbb{R}}
\def \E {\mathbb{E}}
\def \N {\mathbb{N}}
\def \X {\mathcal{X}}
\def \Y {\mathcal{Y}}

\def \F {\mathcal{F}}
\def \EE {\mathcal{E}}

\DeclareMathOperator*{\argmin}{arg\,min}

\def\L{{\cal L}}

\begin{document}

\begin{frontmatter}

\title{DOSED: a deep learning approach to detect multiple sleep micro-events in EEG signal}

\author[stanford,dreem,telecom]{S.~Chambon\corref{cor1}\fnref{fn1}}
\ead{stan.chambon@gmail.com}
\author[dreem]{V.~Thorey\fnref{fn1}}
\ead{valentin@dreem.com}

\author[dreem]{P. J.~Arnal}
\ead{pierrick@dreem.com}
\author[stanford]{E.~Mignot}
\ead{mignot@stanford.edu}
\author[telecom,inria,cea]{A.~Gramfort\corref{cor1}}
\ead{alexandre.gramfort@inria.fr}

\cortext[cor1]{Corresponding authors}
\fntext[fn1]{equally contributed}
\address[stanford]{Center for Sleep Sciences and Medicine, Stanford University, Stanford, California, USA}
\address[dreem]{Research \& Algorithms Team, Dreem, Paris, France}
\address[telecom]{LTCI T\'{e}l\'{e}com ParisTech, Universit\'{e} Paris-Saclay, Paris, France}
\address[inria]{Inria, Universit\'{e} Paris-Saclay, Paris, France}
\address[cea]{CEA Neurospin, Universit\'{e} Paris-Saclay, Paris, France}

\begin{abstract}
%
\textit{Background}: Electroencephalography (EEG) monitors brain activity during sleep and is used to identify sleep disorders. In sleep medicine, clinicians interpret raw EEG signals in so-called sleep stages, which are assigned by experts to every $30$\,s window of signal. For diagnosis, they also rely on shorter prototypical micro-architecture events which exhibit variable durations and shapes, such as spindles, K-complexes or arousals. Annotating such events is traditionally performed by a trained sleep expert, making the process time consuming, tedious and subject to inter-scorer variability. To automate this procedure, various methods have been developed, yet these are event-specific and rely on the extraction of hand-crafted features.

\textit{New method}: We propose a novel deep learning architecure called Dreem One Shot Event Detector (DOSED). DOSED jointly predicts locations, durations and types of events in EEG time series. The proposed approach, applied here on sleep related micro-architecture events, is inspired by object detectors developed for computer vision such as YOLO and SSD. It relies on a convolutional neural network that builds a feature representation from raw EEG signals, as well as two modules performing localization and classification respectively.

\textit{Results and comparison with other methods}: The proposed approach is tested on 4 datasets and 3 types of events (spindles, K-complexes, arousals) and compared to the current state-of-the-art detection algorithms.

\textit{Conclusions}: Results demonstrate the versatility of this new approach and improved performance compared to the current state-of-the-art detection methods.
\end{abstract}


\end{frontmatter}



\section{Introduction}
\label{sec:intro}
Sleep is a behavioral state associated with specific changes in physiology and brain activity patterns~\cite{Porkka-Heiskanen2013}. The most common and practical way to monitor brain activity during sleep is to use electroencephalography (EEG). EEG measures hundreds of times per second the electrical potentials at several locations over the scalp. Identifying in EEG signals microarchitectural events of variable duration like sleep spindles and K-complexes (0.5 - 2\,s duration) or arousals ($\sim10$\,s duration), is of strong interest for sleep research~\cite{iber2007}. Such events are typically used to determine sleep stages, which are scored by 30\,s epochs. In other cases however, when one needs to go beyond sleep scoring, they must be counted and specifically annotated. It is notably the case when the aim is to understand sleep physiology~\cite{Warby2014,Purcell2017} or to study the pathophysiology of specific sleep or neuropsychiatric disorders~\cite{Stephansen2017,Manoach2016,Musiek2015}. The identification of micro-architectural events in the EEG is traditionally performed by trained sleep experts, also called scorers, who visually investigate the recorded signals over a night and annotate the relevant events with their respective start times and durations. This is a tedious, imprecise, costly and time consuming task. Furthermore, this process exhibits a low inter-scorer agreement, which may be improved by taking the consensus of multiple sleep experts~\cite{Warby2014}.

Multiple automatic algorithms for the detection of micro-events in the sleep EEG, such as spindles or K-complexes, have been proposed in the literature. These typically rely on band-pass filtering (typically $11 - 16$\,Hz for spindles, $0.5 - 5$\,Hz for K-complexes), and the extraction of hand-crafted features. Four categories of algorithms can be distinguished.
Methods from the first category extract the envelope of the filtered signal and threshold it~\cite{Ray2015, Wamsley2012, Wendt2012, Moelle2011, Nir2011, Ferrarelli2007}. The thresholding level is then either fixed or tuned. The threshold can be applied to the rectified filtered signal~\cite{Ferrarelli2007}, to the instantaneous amplitude obtained by Hilbert transform~\cite{Nir2011}, to the root mean square of the filtered signal~\cite{Moelle2011}, or alternatively to the moving average of the rectified filtered signal~\cite{Wamsley2012}. In order to identify the start and end times of events, it was for example proposed to look at inflexion points of the envelope of the filtered signal~\cite{Wendt2012}. This process is typically done after signal pre-processing to remove ocular artifacts and environmental noise such as the spurious signals due to electrical current (notch filtering around 50\,Hz or 60\,Hz). One limitation of this first category of methods is that they should be employed during specific sleep stages~\cite{Ferrarelli2007, Nir2011, Moelle2011, Wamsley2012, Ray2015}. Therefore they require preliminary visual inspections of the data and/or a preliminary manual sleep stage scoring.

The second type of approaches decomposes the EEG signals into an oscillatory component and a transient component prior to filtering and thresholding the resulted signals~\cite{Parekh2017, Lajnef2017, Parekh2015}. These methods can detect both spindles and K-complexes, provided some changes are made to hyper-parameters. Also, they are not sleep stage specific which make them more attractive and efficient.
The third type of methods employ unsupervised learning techniques such as clustering. For the clustering step, Patti et al. \citep{Patti2017} use a Gaussian Mixture Model (GMM). First the input signal is a band-pass filtered in 3 frequency bands (10.5 - 16\,Hz, 4 - 10\,Hz, 20 - 40\,Hz), then 2 features called \emph{sigma ratio} and \emph{sigma index} are extracted from sliding windows of $1$\,s: the sigma ratio is the ratio of energy in the spindle frequency band (10.5 - 16\,Hz) during the window of interest over the energy in previous and following windows, while the sigma index is the ratio of power in the spindle band (10.5 - 16\,Hz) over the sum of energy in the neighboring frequency bands (4 - 10\,Hz and 20 - 40\,Hz). The GMM is then applied on extracted features to cluster samples into potential spindles versus non spindles. This approach has the following advantages: it is unsupervised, hence does not rely on human annotations and it is not sleep stage specific. Yet, as the technique is unsupervised the algorithm may not discriminate the events of interest, performance cannot be easily quantified, and the setting of hyper-parameters cannot be automated thanks to cross-validation.
The fourth type of methods are supervised machine learning approaches which consist in training a classifier to predict whether a window of signal is an event of interest or not. For such a binary decision, classifiers such as a Support Vector Machine (SVM) classifier~\cite{Lachner-Piza2018} or a Random Forest Classifier~\cite{Patti2015} can be trained on manually extracted signal features, such as amplitude variance, number of peaks or zero crossings etc.~\cite{Lachner-Piza2018}.

The methods mentioned above suffer from several limitations. First, they rely on pre-defined parameters, such as frequency bands for filtering, which may not be optimal for some recordings or subjects. Second, they are intrinsically event-specific. Third, their hyper-parameters, such as thresholds, are often selected on the recording(s) used for evaluating detection performances, introducing a positive and optimistic bias in reported results.  To address these limitations, possible solutions can be found in the computer vision literature, and more specificaly in the state-of-the-art object detection literature which relies on deep convolutional neural networks~\cite{lin2016fpn,Lin2017,Liu2016, Redmon2015,Ren2015}. Such approaches learn a feature representation that is used by a prediction module that outputs both bounding boxes of detected objects as well as their classes. These approaches can handle objects of multiple classes at any scale and make predictions based on features drawn from entire or subsection of images~\cite{Redmon2015,Ren2015,Lin2017,Liu2016}. Besides, they may make predictions from different features maps of the underlying neural network~\cite{Lin2017,Liu2016} allowing to handle different resolutions and scales of objects. Translating such methods to detect micro-architectural events in EEG time series is of great interest as it would provide the community with a non-event specific and a non sleep stage specific method that predicts locations, durations and types of any \emph{micro-event} at the same time. Nonetheless, the translation from images to EEG time series is not straightforward for two reasons. First, when working on EEG recordings, one needs to process chunks of signals: processing entire nights of signal is not tractable. Second, in a recording, most of these chunks of signals do not contain any true event. This implies that a successful method has to predict not only the absence of any event in the majority of the signal, but start times, durations and classes of events accurately when they occur. In machine learning terms, the method needs to cope with the problem of learning from imbalanced data.
%
%

In this paper we propose the Dreem One Shot Event Detector (DOSED), a deep learning architecture algorithm to detect any type of micro-events in multivariate EEG signals. The proposed approach builds on a convolutional neural network which extracts high-level features from raw non-preprocessed EEG time series. A localization module predicts centers and durations of potential events over the input signals while a classification module predicts their labels. The whole network architecture is trained end-to-end by back-propagation. In the following sections, we first detail this general approach and the associated training procedure. We then present a detailed and extensive benchmark comparing the proposed method with multiple state-of-the-art algorithms, on $3$ event detection tasks and over $4$ datasets. We also address technical questions pertaining to the influences of hyper-parameters on detection performances. This work extends on a recent short communication~\cite{Chambon2018a}.


\section{Methods}
\paragraph*{Notation}
We denote by $\llbracket n \rrbracket$ the set of integers $\{1, \ldots, n\}$ for $n \in \N$. Let $\X = \R^{C \times T}$ be the set of EEG input samples where $C$ stands for the number of EEG channels and $T$ for the number of time steps. Let $L \in \N$ be the number of different labels or types of events to be predicted. We denote $\L = \llbracket L \rrbracket$ the set of events labels, and $0$ the label associated to no event or background signal. An event $e = \left\{t^c, t^d, l \right\} \in \EE = \R^2 \times \L \cup \{0 \}$ is defined by a center location time $t^c$, a duration $t^d$ and an event label (or type) $l \in \L  \cup \{0\}$. A \emph{true event} is an event with label in $\L$ detected by a human scorer or a group of human scorers, \emph{a.k.a.} \emph{consensus}. A \emph{predicted event} is an event with label in $\L$ detected by an algorithm or a group of algorithms.
%
%
\subsection{Method overview}
\begin{figure}[ht!]
\centering
\includegraphics[width=0.98\linewidth]{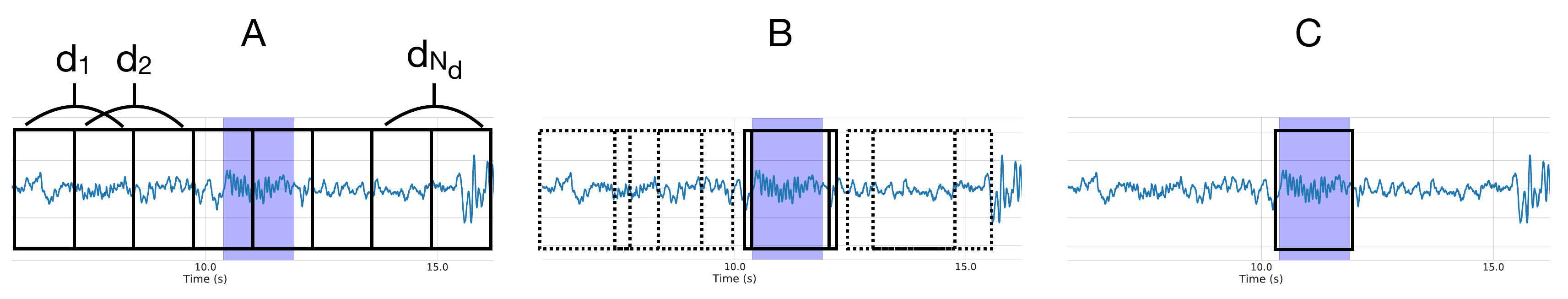}
\caption{\label{fig:scheme}Prediction procedure for DOSED. A: $N_d$ default events, $d_i$, $i \in \llbracket N_d \rrbracket$, are generated over the EEG sample. B: the network predicts potential events, \emph{i.e.} adjusted centers and durations with respect to default events centers and durations, and potential events labels. C: non-maximum suppression is applied to merge overlapping potential events with label $l$ different from $0$. The network finally returns the center(s), duration(s) and label(s) of the remaining merged event(s).
}
\end{figure}

The general principle of DOSED is as follows. Let $x \in \X$ be an EEG sample, \emph{i.e.} a short window of signal ($\sim$ 30\,s duration) coming from a PSG recording ($\sim$ 8\,h). First, the method is initialized with $N_d$ default events $d_i = (t_i^c, t_i^d)$, which are parameterized with a center time $t_i^c$ and a duration $t_i^d$, and which are positioned over each input signal $x$. For example, 1\,s default events every 0.5\,s if this corresponds to a typical duration of events to be detected, see Figure~\ref{fig:scheme}~-~A. Note that, centers, durations and overlaps of such default events can be adjusted depending on the type(s) of event(s) to be detected. Second, the network predicts a potential event associated to each default event: it predicts an adjusted center and an adjusted duration, as well as the probability for this event to have any label $l \in \L \cup \{ 0 \}$, cf. Figure~\ref{fig:scheme}~-~B. Finally, potential events
for which the highest probability label $l \in \L$, and for which the probability is higher than a specific (cross-validated) threshold $\theta_l \in [0, 1]$, are selected. Then, non-maximum suppression is applied to remove overlapping events, cf. Figure~\ref{fig:scheme}~-~C~\cite{Liu2016,Redmon2015}.
%
%

\subsection{Loss function}

Supervised machine learning boils down to the minimization of a specific loss function that quantifies the prediction errors from the model on the training database.
Here the default events are parameterized by a center and a duration, so the method predicts an adjusted center and an adjusted duration for any default event. It also returns the probabilities of each event type as well as the probability of belonging to the background signal ($l=0$). 

To quantify the overlap in time between two events, a commonly used metric is the Jaccard index, \emph{a.k.a.} Intersection over Union (IoU)~\cite{Redmon2015}. For two time intervals, it is defined as the ratio of size between their intersection and their union. This metric takes values between zero and one. It is zero if the events do not overlap in time, and one if they perfectly overlap.

The goal is to learn a prediction function $\hat{f}$ from $\X$ to $\Y$ where $y \in \Y$ is a set of elements from $\EE$. Given $N_d \in \N$ default events generated over the input EEG sample $x \in \X$. Let $D(x) = \left\{ d_i = (t^c_i, t^d_i), i \in \llbracket N_d \rrbracket \right\}$ be the set of centers and durations of the $N_d$ default events generated over $x$. Let $E(x) = \left\{ e_j = (t^c_j, t^d_j, l_j) : j \in \llbracket N_e \rrbracket \right\}$ be the list of the $N_e$ true events annotated over the signal $x$. 

At training time, we want to map the default events and the true events. Following SSD~\cite{Liu2016} we use per-prediction matching. First, bipartite matching is applied: each true event is matched with the default event presenting the highest IoU. The remaining default events are then matched to the ground truth event presenting the highest IoU with $\mathrm{IoU}(d_i, e_j) > \eta \in [0,1]$. We denote by $\gamma$ the function which returns, if it exists, the index $j$ of the true event matching with the default event $d_i$, and $\emptyset$ otherwise. The unmatched default events are assigned with the label $l=0$ related to the absence of event.

Let $e_j$ be a true event matching with the default event $d_i$. $d_i$'s center and duration are then encoded with $\phi_{e_j}: \R^2 \longrightarrow \R^2, d_i = (t^c_i, t^d_i) \longmapsto \left( \dfrac{t^c_j - t^c_i}{t^d_i}, \log \dfrac{t^d_j}{t^d_i} \right)$~\cite{Ren2015}. This encoding function quantifies the relative variations in centers and durations between the default event $d_i$ and the true event $e_j$.

Let $\hat{f}(x) \in \Y$ be the prediction made by the model $\hat{f}$ over the sample $x$. We define it as $\hat{f}(x) = \{(\hat{t}^c_i, \hat{t}^d_i, \hat{l}_i) \in \EE, i \in \llbracket N_d \rrbracket \}$. $\hat{\tau}_i = (\hat{t}^c_i, \hat{t}^d_i)$ are the predicted coordinates of encoded default event $d_i$ and $\hat{l}_i$ is its predicted label. In practice, the model will output the probability of each label $l \in \L \cup \{ 0 \}$ for default event $d_i$ so $\hat{l}_i$ is replaced by a vector of probabilities $\hat{\pi}_i \in [0, 1]^{ |\L| + 1}$. As it is a probability vector, we have $\sum_{l \in \L \cup  \{0 \}} \hat{\pi}_i^l = 1$.

The loss between the true annotation $E(x)$ and the model prediction $\hat{f}(x)$ over signal $x$ is
a function $\ell: \Y \times \Y \rightarrow \R_+$ defined as $\ell \left( E(x), \hat{f}(x) \right) = \ell^+_{norm} + \ell^-_{norm}$ where
\begin{align}
\ell^+ &= \sum_{\substack{ i \in \llbracket N_d \rrbracket \\ \gamma(i) \neq \emptyset }} \mathrm{L1}_{smooth} \left( \phi_{ e_{\gamma (i)} }(d_i) - \hat{\tau}_i \right) - \log(\hat{\pi}_i^{l_{\gamma (i)}}) \label{eq:loc_clf_pos}\\
\ell^- &= - \sum_{\substack{i \in \llbracket N_d \rrbracket \\
                 \gamma(i) = \emptyset }}
            \log(\hat{\pi}_i^0)\label{eq:clf_neg}
\end{align}
and the normalized losses $\ell^{+}_{norm}$ and $\ell^-_{norm}$ are obtained by dividing $\ell^{+}$ and $\ell^-$ by the numbers of terms involved in the sums~\eqref{eq:loc_clf_pos} and~\eqref{eq:clf_neg} respectively. In \eqref{eq:loc_clf_pos}, the sum considers the localization and classification losses for any default event $d_i$ matching a true event $e_{\gamma(i)}$. The $\mathrm{L1}_{smooth}$ loss applies coordinate-wise the real valued function: $x \mapsto (x^2 / 2) \mathds{1}_{|x| < 1} + (|x| - 1/2) \mathds{1}_{|x| \geq 1}$, a.k.a. the Huber loss~\cite{Ren2015}. The purpose of~\eqref{eq:loc_clf_pos} is to promote accurate localization and classification on default events matching a true event ($l \in \L$). Equation~\eqref{eq:clf_neg} corresponds to the classification loss for default events which do not match any true event ($l = 0$).

One particular difficulty for the proposed approach, is that most of the default events do not match any true event. This is due to the facts that (1) EEG signals contain few events over long time periods, and that (2) the parameterization of default events implies using as many overlapping default events as necessary to maximize the number of true events matched over a window $x$. To address the issue of label imbalance between default events matching a true event and default events not matching any true event ($l=0$), we employed a negative-mining step: during training only a fraction of default events with label $l = 0$ were used. Let us remark that this negative mining step is omitted in \eqref{eq:clf_neg} for simplicity. The process first boils down to sorting default events which do not match any true event based on their classification errors. The events which exhibit the worst classification errors are then considered, and a fraction of them is finally selected to compute the loss \eqref{eq:clf_neg}. This fraction is set such that the ratio between default events matching a true event and default events matching no true event is equal to $1 / 3$, while ensuring that a minimum number of $10$ default events matching no true event is selected.

Remark that the parameterization of default events is important to ensure a successful training process. In particular the parameterization of the duration $t_i^d$ of the default events is critical. Indeed, a bad choice for $t_i^d$ might lead to a small number of matched true events during the training process thus hurting the detection performances of the proposed approach. The influence of the default events durations is investigated and discussed in a specific experiment
in Section~\ref{sec:expes}.

To conclude, training the event detection method $\hat{f}$ boils down to solving the following minimization problem:
\begin{equation}\label{eq:training_objective}
\hat{f} \in \argmin_{f \in \F} \E_{x \in \X} \left[ \ell \left(E(x), f(x) \right) \right]
\end{equation}

Note that the loss function $\ell$ is differentiable with respect to the parameters of the considered model $f$, so it becomes possible to learn a neural network $f$ with (stochastic) gradient methods and back-propagation. Also note that in practice, one minimizes the empirical version of~\eqref{eq:training_objective} computed over a finite number of labeled training samples.

\subsection{Model architecture}
What remains to be defined is the form of the function $f$, i.e. the architecture of the network.
We use for $f$ a deep convolutional neural network that predicts at most $N_d$ potential events with label different from $0$ given a set of default events $D(x) = \{d_i = (t_i^c, \ t_i^d): i \in \llbracket N_d \rrbracket \}$.

The architecture of the network is as follows. First, following the work in~\cite{Chambon2018a,Chambon2018}, it starts with a spatial filtering of the input signals. This corresponds to a matrix multiplication similar to Independent Component Analysis (ICA) that increases the signal to noise ratio~\cite{Chambon2018a, Chambon2018}. Then, it performs some temporal processing with convolutions to build a feature representation of the input signals, and finally it outputs the prediction of intervals and labels of potential events.

Let $K, F, C, T \in \N$. The network can be written as the composition of three functions, $f(x) = \psi(\phi_T(\phi_C(x)))$. The module $\phi_C: \X \longrightarrow \X$ stands for the spatial filtering operation which performs $C$ linear combinations of the $C$ input time series and returns a new tensor $\tilde{x} \in \X$. This module is implemented by a 2D convolution layer with $C$ kernels of size $(C, 1)$ (space, time) followed by a transpose operation. When $C=1$ this module becomes the identity function: $\phi_C = \mathrm{Id}$.

The module $\phi_T: \X \longrightarrow \R^{F \times C \times \tilde{T}}$ performs the temporal feature extraction. The module $\phi_T$ is composed of $K$ blocks. Each block $k$ is composed of a 2D convolution layer with batch normalization~\cite{Ioffe2015} and ReLU activation $x \mapsto \max(x, 0)$~\cite{Nair2010}, followed by a temporal max-pooling. Block $k$ first convolves the previous feature maps $x_{k-1}$ with $4 \times 2^k$ kernels of size $(1, 3)$ (space, time), using a stride of $1$. Zero padding is used to maintain the dimension of the tensor through the convolution layer. Then, the ReLU activation is applied. Finally a temporal max pooling operation with kernel of size $(1, 2)$ and stride $2$ is applied to divide by 2 the temporal dimension. Each Block $k$ does not process the spatial dimension. Thus, the output of $\phi_T$ is a tensor of shape $(F, C, \tilde{T})$ with $F = 4 \times 2^K$ and $\tilde{T} = T / 2^K$.

The module $\psi: \R^{F \times C \times \tilde{T}} \longleftrightarrow (\R^2 \times \L \cup \{0 \})^{N_d}$, stands for the prediction part of the model: it predicts for each of the $N_d$ default events its corresponding potential event, \emph{i.e.} its adjusted center and duration in $\R^2$ as well as the label of this potential event in $\L \cup \{ 0 \}$. In practice it predicts a probability vector which encodes the probability of that event to belong to any of the classes $\L \cup \{ 0 \}$. The prediction of locations is implemented by a convolution layer with $2 \times N_d$ kernels of dimension $(C, \tilde{T})$ and a linear activation. The prediction of classes is implemented with a 2D convolution layer with $(L + 1) \times N_d$ kernels of dimension $(C, \tilde{T})$. A softmax activation is applied on every $L + 1$ output feature maps to obtain a probability vector $\pi_i$ for any potential event $i$. The general architecture is summarized in Table~\ref{tab:phi}.

\begin{table*}[ht!]
\centering
\begin{tabular}{c|llllll}
Module                                                                                                                                   & Layer Type & \begin{tabular}[c]{@{}l@{}}kernel\\ size\end{tabular} & \begin{tabular}[c]{@{}l@{}}kernel\\ \#\end{tabular} & output dim                       & activation                                                              & stride   \\ \hline
\rowcolor[HTML]{EFEFEF} 
\cellcolor[HTML]{EFEFEF}                                                                                                                 & Conv. 2D   & $(C, 1)$                                              & $C$                                                 & $(C, 1, T)$                      & linear                                                                  & 1        \\
\rowcolor[HTML]{EFEFEF} 
\multirow{-2}{*}{\cellcolor[HTML]{EFEFEF}$\phi_C$}                                                                                       & Transpose  & -                                                     & -                                                   & $(1, C, T)$                      & -                                                                       & -        \\ \hline
\rowcolor[HTML]{EFEFEF} 
\cellcolor[HTML]{EFEFEF}                                                                                                                 & Conv. 2D   & $(1, 3)$                                              & $4 \times 2^k$                                      & $(4 \times 2^k, C, T / 2^{k-1})$ & relu                                                                    & 1        \\
\rowcolor[HTML]{EFEFEF} 
\multirow{-2}{*}{\cellcolor[HTML]{EFEFEF}\begin{tabular}[c]{@{}c@{}}$\phi_T$\\ $k$ blocks, $k \in \llbracket K \rrbracket$\end{tabular}} & Max P. 2D  & $(1, 2)$                                              & -                                                   & $(4 \times 2^k, C, T / 2^k)$     & -                                                                       & $(1, 2)$ \\ \hline
\rowcolor[HTML]{C0C0C0} 
$\psi$ - localization                                                                                                                    & Conv. 2D   & $(C, \tilde{T})$                                      & $2 \times N_d$                                      & $(2 \times N_d, 1, 1)$           & linear                                                                  &          \\
\rowcolor[HTML]{C0C0C0} 
$\psi$ - classification                                                                                                                  & Conv. 2D   & $(C, \tilde{T})$                                      & $(L + 1) \times N_d$                                & $((L + 1) \times N_d, 1, 1)$     & \begin{tabular}[c]{@{}l@{}}softmax\\ every (L + 1) kernels\end{tabular} &         
\end{tabular}
\caption{General architecture of the proposed approach. The network $f(x) = \psi(\phi_T(\phi_C(x)))$ processes an input tensor of shape $(1, C, T)$ corresponding to $C$ time series of $T$ time steps. $\phi_C$ stands for the spatial processing module that performs spatial filtering. When $C=1$, $\phi_C = \mathrm{Id}$. $\phi_T$ stands for the temporal processing module that builds the temporal feature representation and outputs a tensor of shape $(F, C, \tilde{T})$ with $F=4 \times 2^K$ and $\tilde{T} = T / 2^{K}$. $\psi$ stands for the prediction module that predicts locations of potential events and their classes, \emph{i.e.} it outputs two tensors: the localization tensor of shape $(2 \times N_d)$ and the classification tensor of shape $((L + 1) \times N_d)$. Note that the predictions exploit the full spatio-temporal data: they are based on convolutions of the whole output of $\phi_T$}
\label{tab:phi}
\end{table*}

\section{Experiments}
\label{sec:expes}

\subsection{Datasets}

The experiments were performed on 4 datasets described below. Details on the datasets are summarized in Table~\ref{tab:dataset_properties}.

\paragraph*{Montreal Archives of Sleep Studies dataset}
The publicly available dataset Montreal Archives of Sleep Studies - session 2 (MASS SS2)~\cite{OReilly2014} was used. This dataset contains 19 polysomnography (PSG) recordings (11 females, 8 males, 23.6 $\pm$ 3.7 years old). These were sampled at $256$\,Hz. Spindles and K-complexes have been manually annotated by a first expert (E1) over $19$ PSGs, and spindles have been annotated by a second expert (E2) over $15$ PSGs using different guidelines~\cite{Lajnef2017}. All annotations have been performed over the EEG channel C3. In this work, only annotations produced by scorer E1 and data from channel C3 were used.

\paragraph*{Stanford Sleep Cohort dataset}
The second dataset includes $26$ PSGs from the patient-based Stanford Sleep Cohort (SSC), see~\cite{Andlauer2013} for a description of this dataset. These PSGs are from $26$ subjects (9 females, 17 males, 52.2 $\pm$ 14.3 years old) and were sampled at $128$\,Hz. For each PSG recording, spindles have been scored by $5$ different sleep experts through visual investigation of central channels. Experts were randomly selected from a pool of $9$ scorers. For each PSG recording, annotations made by $5$ experts were merged to build a \emph{consensus}. To do so, annotations of each expert $j$ were first encoded as binary vectors $y_j \in \{0, 1\}^T$ containing as many time steps $T$ as in the considered PSG recording. A time step $i$ was considered as a spindle, $y_j[i] = 1$, if it belonged to a spindle annotated by the expert $j$, otherwise it was considered as background signal and $y_j[i] = 0$. The binary vectors corresponding to the five annotations were then averaged to build a vector $\bar{y} \in [0, 1]^T$, taking values in $\{0, 0.2, 0.4, 0.6, 0.8, 1\}$. The consensus annotations were obtained by thresholding the vector $\bar{y}$ to get a binary vector. Let $\kappa \in [0, 1]$ stand for the desired level of consensus. The binary vector $y_{\kappa} \in \{0, 1\}^T$ was built as follows.
\begin{equation*}
y_{\kappa} [i] = \left\{
    \begin{array}{ll}
        1 & \mbox{if } \bar{y} \geq \kappa\\
        0 & \mbox{otherwise }
    \end{array}
\right.
\forall i \in \llbracket T \rrbracket
\end{equation*}
Start times and durations for all events were then estimated from $y_{\kappa}$.

Unless otherwise mentioned, we used the consensus $\kappa=0.2$ (union of scorers) annotations from channel C4 (C3 if C4 not available), using as reference the mastoid electrode.

\paragraph*{Wisconsin Sleep Cohort dataset}
Third, $30$ PSG recordings from $30$ subjects of the Wisconsin Sleep Cohort (WSC), were considered~\cite{Young2008} (16 females, 14 males, 65.2 $\pm$ 8.2 years old). The PSGs were sampled at $200$\,Hz. Similarly to SSC, each PSG was scored by $5$ different sleep experts from the same pool of $9$ possible scorers. All experts followed the same scoring guidelines. Unless otherwise mentioned, we used consensus $\kappa=0.2$ (union of scorers) annotations on signals from channel C4 (C3 if C4 not available), using as reference the mastoid electrode.

One reason for using a consensus vector on SSC and WSC is the quality of some annotations. Indeed some annotations are questionable as certain experts were only able to annotate start times and durations up to a $0.5$\,s precision.

\paragraph*{MESA dataset}
A fourth dataset, publicly available, was finally used: MESA~\cite{Dean2016}. $1000$ PSG recordings from $1000$ subjects (528 females, 472 males, 69.3 $\pm$ 9.0 years old) originally sampled at $256$\,Hz were used for the experiments. The signals from the 3 EEG channels (Fz-Cz, Cz-Oz, C4), and the 2 EOG channels (EOG left and right) were considered in this work. MESA dataset was used to study the detection of arousals which have been annotated based on visual investigation of C4 channel.


\subsection{Evaluation methodology}

\paragraph*{Cross-validation}
A $5$ split cross-validation was used on SS2, SSC and WSC. On SS2, a split stands for $10$ training, $5$ validation and $4$ testing PSG recordings. On SSC, a split corresponds to $15$ training, $5$ validation and $6$ testing PSGs. For WSC, a split stands for $19$ training, $5$ validation and $6$ testing PSGs. On MESA, a single split was performed, using $400$ PSGs for training, $100$ for validation and $500$ for testing. Cross-validation details are summarized in Table~\ref{tab:dataset_properties}.


\paragraph*{Metrics}
\emph{By event metrics} were used to quantify detector performances for detection and localization  of events~\cite{Warby2014}. These metric rely on the following $\mathrm{IoU}$ criterion: for a given $\delta > 0$, a predicted event was considered as a true positive if it exhibited an $\mathrm{IoU} \geq \delta$ with a true event, otherwise it was considered as a false positive. The numbers of positives and true positives were evaluated to compute precision, recall and  F1 scores of detectors for different overlapping criterion $\delta \in \{0.1, 0.2, \dots, 0.9 \}$. When $\delta = 0.1$ a predicted event which exhibits at least some small overlap with a true event might be considered a true positive, and when $\delta = 0.9$ only a predicted event which exhibits a high overlap with a true event is considered a true positive. Evaluation was performed for $\delta \in \{0.1, 0.2, \dots, 0.9 \}$, on entire PSGs, each being taken individually. Reported performances were averaged by values of $\delta$ over PSG recordings from the testing set.

\paragraph*{Compared methods}
For spindles detection, $3$ state-of-the-art alternative methods were compared: \emph{Parekh et al. 2017}~\cite{Parekh2017}, \emph{Lajnef et al. 2017}~\cite{Lajnef2017} and \emph{Lachner-Piza et al. 2018}~\cite{Lachner-Piza2018}. Benchmarks relied on codes provided by the original authors\footnote{\emph{Lajnef et al. 2017}: \url{https://github.com/TarekLaj/SPINKY}, \emph{Parekh et al. 2017}: \url{https://github.com/aparek/mcsleep}, \emph{Lachner-Piza et al. 2018}: \url{https://github.com/mossdet/Mossdet}}. For K-complexes detection, \emph{Lajnef et al. 2017}~\cite{Lajnef2017} was considered as the baseline comparator. This algorithm was designed to detect negative peaks of K-complexes without detecting precisely start times and end times of these events. The convention employed by the authors was therefore used: start time was predicted as $0.1$\,s before the negative peak and end time as $1.3$\,s after the negative peak. Hyper-parameters of \emph{Parekh et al. 2017}, $\lambda_3$ and $threshold$ were searched over $\{10, 20, 30, 40, 50\} \times \{0.5, 1, 1.5, 2, 2.5, 3\}$. Threshold parameter of \emph{Lajnef et al. 2017} was selected in $\{0, 25, \dots, 250\}$ for spindles detection and in $\{-100, -95, -90, \dots, 0\}$ for K-complexes detection.

Hyper-parameters were selected by grid search on the training and validation PSG recordings at hand. The selection was performed in order to maximize F1 scores. More precisely, for a given overlapping criterion $\delta \in \{0.1, \dots, 0.9 \}$ of interest, F1-scores associated to every set of hyper-parameters were evaluated on validation PSG recordings. The set achieving the highest F1-score was selected to predict PSG metrics of the testing set, computed with respect to this specific $\delta$. The process is repeated for every $\delta \in \{0.1, 0.2, \dots, 0.9 \}$, meaning that for each $\delta$ a potentially different set of optimal hyper-parameters is selected.

For \emph{Lacher-Piza et al. 2018}, the code provided by the authors came as a stand-alone software allowing neither retraining of the underlying detector nor hyper-parameter tuning. We therefore asked for \emph{Lachner-Piza et al. 2018}' detector to perform well across unknown datasets without hyper-parameter selection, a more difficult task referred to as inter dataset generalization. This likely led to lower performances than the ones this approach could potentially deliver if training and hyper-parameters tuning had been possible.

\paragraph*{Proposed approach}
For detection of spindles and K-complexes, the network was provided with $20$\,s EEG samples, sampled at a sampling frequency $F_s$ specific to each dataset. Furthermore, the feature extraction module $\phi_T$ was built with $K=8$ blocks. Thus, $x \in \R^{C \times T}$, with $T = 20 \times F_s$. Only the case $C = 1$ was considered for spindles and K-complexes. For arousals detection, the network was provided with $2$\,min samples, down-sampled to $128$\,Hz. For this task, if not mentioned, $C = 1$. For every task, a normalization was applied to each sample $x$: centering and standardization by dividing each centered signal by its standard deviation computed on the full recording. We summarize the training details of the proposed approach on each dataset in Table~\ref{tab:dataset_properties}.

\begin{table}[ht!]
\centering
\begin{tabular}{l|llllll}
\rowcolor[HTML]{C0C0C0} 
{\color[HTML]{000000} } & {\color[HTML]{000000} $F_s$ (Hz)} & {\color[HTML]{000000} Duration (s)} & {\color[HTML]{000000} T} & {\color[HTML]{000000} C} & {\color[HTML]{000000} $\mathrm{lr}$} & {\color[HTML]{000000} \begin{tabular}[c]{@{}l@{}}\# records\\ (training / validation / testing)\end{tabular}} \\ \hline
SS2                     & 256                               & 20                                  & 5120                     & 1                        & $10^{-4}$                            & 10 / 5 / 4                                                                                       \\
SSC                     & 128                               & 20                                  & 2560                     & 1                        & $10^{-3}$                            & 15 / 5 / 6                                                                                       \\
WSC                     & 200                               & 20                                  & 4000                     & 1                        & $10^{-3}$                            & 19 / 5 / 5                                                                                       \\
MESA                    & $128^*$                               & 120                                 & 15360                    & 1 - 5                    & $10^{-3}$                            & 400 / 100 / 500
\end{tabular}
\caption{Datasets properties and training parameters we considered per dataset. $F_s$ is the sampling frequency of the given dataset. Duration stands for the duration of the input windows used in the the proposed approach. T is the resulting number of time steps. C stands for the number of channels that was considered. $\mathrm{lr}$ stands for the learning rate used during stochastic gradient descent training. $^*$: Note that PSG recordings from MESA were down-sampled to $128$\,Hz, which is in practice performed at the level of batch-sampling}
\label{tab:dataset_properties}
\end{table}

This approach was implemented using the PyTorch library~\cite{paszke2017automatic}. Minimizing~\eqref{eq:training_objective} was achieved using a stochastic gradient descent, with a learning rate of $\mathrm{lr} = 10^{-4}$ on MASS ($\mathrm{lr} = 10^{-3}$ on SSC, WSC, MESA), a momentum $\mu = 0.9$ and a batch size of $32$. $200$ training epochs were considered. If not explicitly mentioned, each batch was balanced \emph{i.e.}, containing $50\%$ of samples with at least one true event and $50\%$ of samples without any true event. In practice, samples were drawn at a random time location in the record until they match the condition of having at least one event or not. When a true event was partially included in a sample, its label $l$ was set to $0$ if less than $50\%$ of this event was part of that sample. Early stopping was used to stop the training process when no improvement was observed on the loss evaluated on the validation data over $10$ consecutive epochs. Furthermore, a learning rate decrease procedure was used: when no progress was observed on the validation loss after $5$ consecutive epochs, the learning rate was divided by $2$. Learning rate decrease on plateau is commonly used for training neural networks~\cite{Goodfellow-et-al-2016}. At prediction time, consecutive EEG samples were sampled from entire PSG recordings, and the network predicted on each of these samples.

Matching hyper-parameter $\eta$ was fixed to $\eta = 0.5$. The proportion of default events containing a true event versus the defaults events containing no true event was fixed to $1 / 3$, and the minimum number of default events containing no true event was fixed to $10$. This ensures that class unbalance between default events matching a true event and those which do not is limited to $1 / 3$ which is assumed to be better for the classification module of the network. Selecting a minimal number of default events matching no true event ensures that the network learns something on each sample, even if it does not contain a true event. Non-maximum suppression was applied to merge potential events exhibiting at least an $\mathrm{IoU} \geq 0.4$.
Default events for spindles and K-complexes detection were fixed as 1\,s sliding windows with 75\% overlap between two consecutive default events, resulting in $N_d=80$ default events.
Default event hyper-parameters for arousals detection were investigated in a dedicated experiment. A potential event $(\hat{t}_i^c, \hat{t}_i^d, \hat{\pi}_i)$ was considered a positive event of label $l$ if $\hat{\pi}_i^l \geq \theta^l$, $\theta^l \in [0, 1]$ being a detection threshold specific to label $l$. This means that a predicted event was considered as a positive event of label $l \in \L$ if the probability of this label $\hat{\pi}^l_i$ is higher than a certain cross-validated detection threshold $\theta^l$.

The detection threshold $\theta^l$ for detecting events of label $l \in \L$ was selected by cross-validation for any overlapping criterion of interest $\delta \in \{ 0.1, \dots, 0.9 \}$ following a process similar to the one used for the baselines. For every $\delta$, hyper-parameter $\theta^l$ was selected by grid search over the validation data to maximize the F1 score. More precisely, for a given $\delta$, the network was used to predict on the validation PSG recordings with different detection thresholds $\theta^l \in [0, 1]$ leading to different precision, recall and F1 scores. The detection threshold achieving the highest F1 score with respect to this criterion $\delta$ was used for performance evaluation on the testing set. The process was repeated for any $\delta \in \{0.1, \dots, 0.9\}$.

In summary, to use the proposed approach one needs to select the following parameters: (1) the default events parameters depending on the events to detect (\emph{a priori} knowledge), (2) the learning rate by monitoring the training and validation losses (\emph{experimental knowledge}) (3) the detection threshold $\theta^l$ by cross-validation (\emph{experimental knowledge}).

%
%

\subsection{Results}

We now provide the results on the different detection tasks (spindles, K-complexes, arousals).

\paragraph*{Spindles detection}

\begin{figure*}[ht!]
\centering
\includegraphics[width=\linewidth]{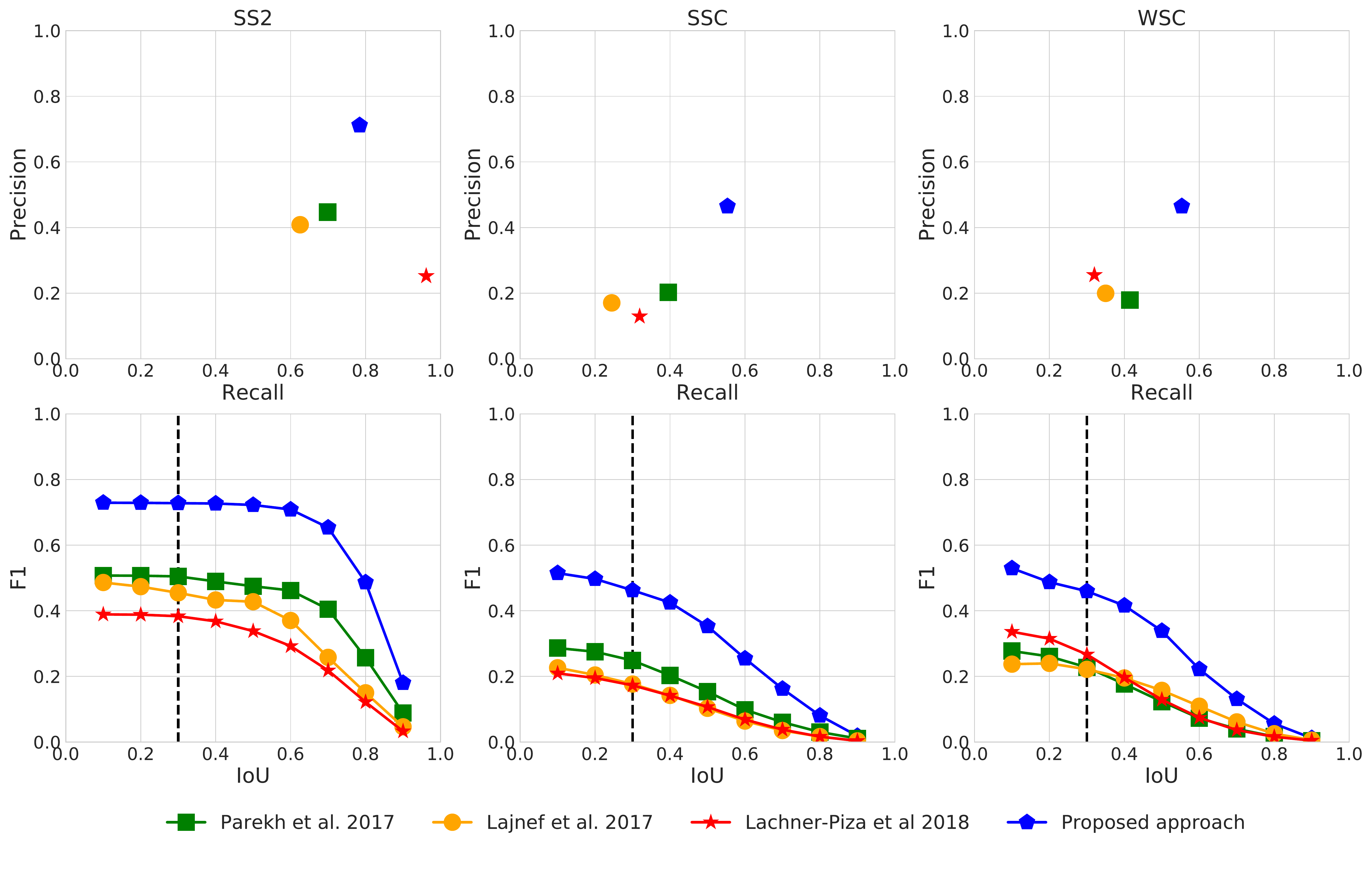}
\caption{Spindle detection: general benchmark on 3 different datasets: SS2, SSC and WSC. First row: averaged precision / recall at $\mathrm{IoU} = 0.3$. Second row: F1 score as a function of IoU. Standard deviation at each IoU are provided as errorbars. The proposed approach outperforms the 3 alternative methods on each metric (Precision, recall and F1 score).}
\label{fig:benchmark_spindles_averaged}
\end{figure*}

\begin{figure*}[ht!]
\centering
\includegraphics[width=\linewidth]{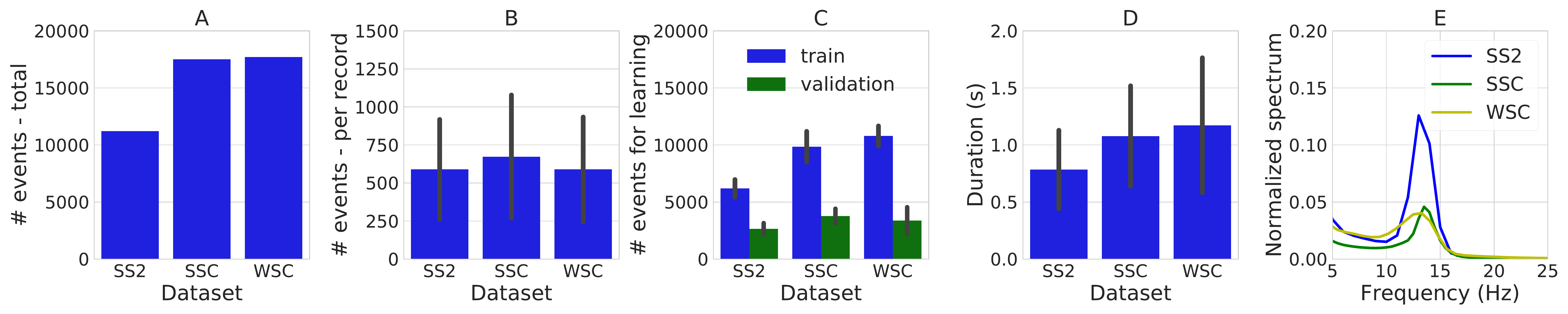}
\caption{Statistics about spindles over the different datasets: the datasets contain comparable amounts of labeled events, yet the frequency content corresponding to spindles $11 - 16$\,Hz is more salient on SS2 than on SSC / WSC.  A: total number of spindles. B: averaged number of spindles per PSG recording. C: averaged numbers of events used for training and validation. D: averaged durations of events. E: normalized spectrum of $2$\,s of signal centered on spindles. This is actually the spectrum averaged over all the spindles annotated on a dataset. The normalization is obtained by dividing the power of each frequency bin by the total power between $[0, F_s / 2]$}
\label{fig:statistics_spindles}
%
%
\end{figure*}

In this experiment, we compare the proposed approach with three alternative methods on the spindle detection task over 3 different datasets: SS2, SSC and WSC. We perform $3$ intra-dataset benchmarks. We report the detection performances of every approach obtained on each dataset in Figure~\ref{fig:benchmark_spindles_averaged}. We also report some statistics about the datasets in Figure~\ref{fig:statistics_spindles}.

A first observation is that, on every dataset, the proposed approach outperforms the compared methods in terms of precision / recall at IoU = 0.3 and in terms of F1 score for any IoU. One can also see on SS2, that the detection performances are quite stable for any IoU in $\left[0.1, 0.7\right]$.

A second major observation is that the learning task appears more difficult on SSC and WSC datasets than on SS2. Indeed, performances reported on SS2 are higher than those reported on SSC and WSC. Also, the standard deviations obtained on SS2 are smaller than those obtained on SSC and WSC. Finally the losses obtained on SS2 are smaller than the ones obtained on SSC and WSC, see Figure~\ref{fig:benchmark_spindles_sup} and Figure~\ref{fig:learning_losses} in Appendix. This observation is made, although the statistics about the datasets are quite similar, especially regarding the quantity of training samples, see Figure~\ref{fig:statistics_spindles}. Yet, the frequency analysis indicates that the spindles from SS2 exhibit a more salient frequency content in the $11 - 16$\,Hz frequency band compared to SSC / WSC, see Figure~\ref{fig:statistics_spindles} - E. This might explain the lower performances obtained on SSC and WSC compared to the ones obtained on SS2.

%
%

\begin{figure*}[ht!]
\centering
\includegraphics[width=\linewidth]{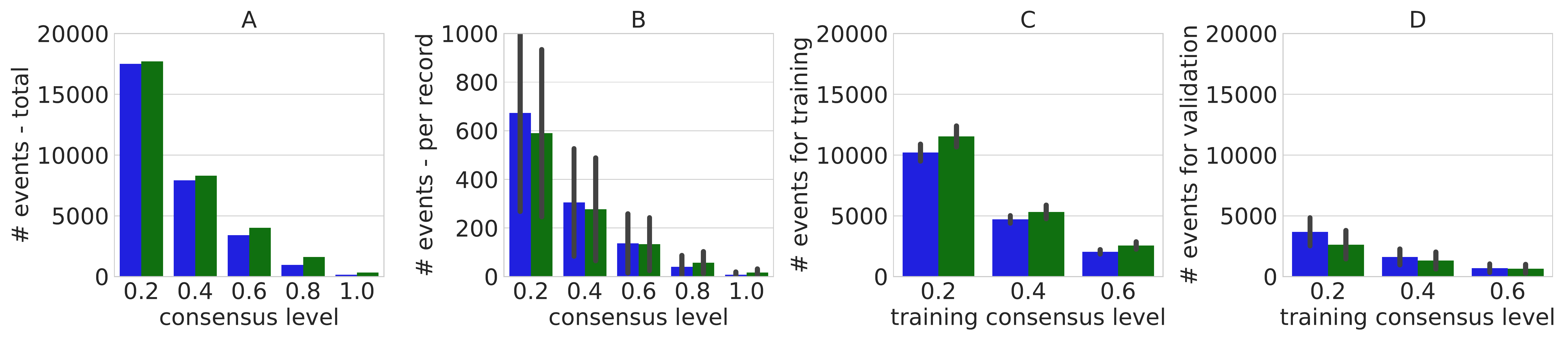}
\caption{Statistics about spindles on SSC and WSC as functions of the consensus level: the consensus level influences greatly the number and the duration of the events annotated by the consensus of scorers.  A: total number of spindles. B: averaged number of spindles per PSG recording. C: averaged numbers of events used for training. D: averaged numbers of events used for validation}
\label{fig:statistics_spindles_consensus}
\end{figure*}

\begin{figure*}[ht!]
\centering
\includegraphics[width=0.73\linewidth]{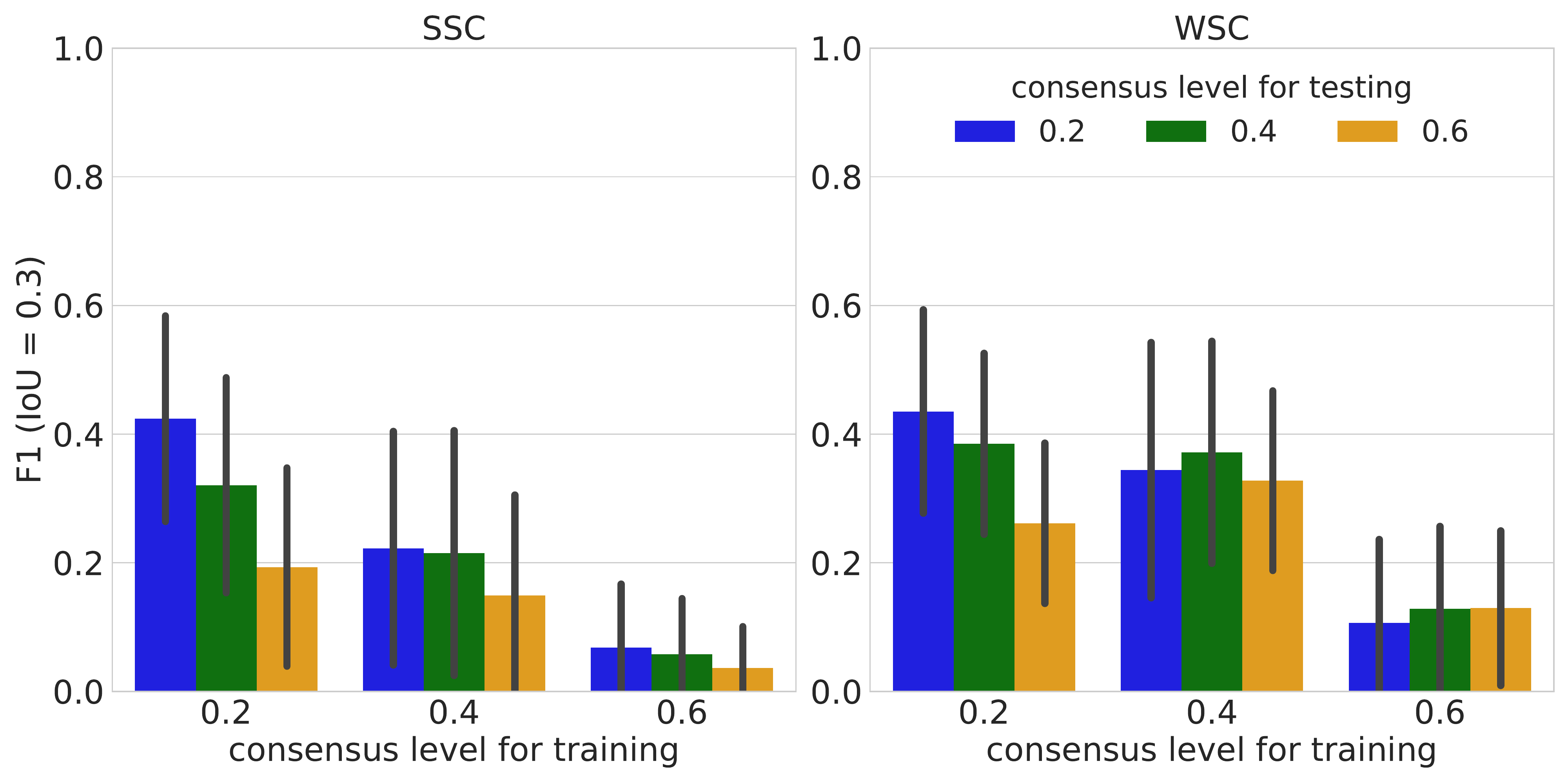}
\caption{Benchmark on spindles detection as a function of the consensus levels used for training and testing on WSC and SSC: the F1 score at IoU = 0.3 is greatly impacted by both the consensus level used for training and the consensus level used for testing}
\label{fig:benchmark_spindles_consensus}
\end{figure*}

\paragraph*{Spindle detection and consensus level}
In this experiment, we benchmark the proposed approach on the spindles detection task, on SSC and WSC datasets. Annotations built with $3$ different consensus levels are considered: $\kappa \in \{0.2, 0.4, 0.6\}$. We report in Figure~\ref{fig:statistics_spindles_consensus} the statistics of annotated events depending on the consensus level considered. The obtained performances are reported in Figure~\ref{fig:benchmark_spindles_consensus}.

%
%

One can first observe in Figure~\ref{fig:statistics_spindles_consensus} that the numbers and the durations of events decrease as the consensus level increases. This leads to fewer training and validation events for learning when $\kappa$ increases. The changes induced by the consensus level on the annotated events statistics induce significant changes in the obtained performances, see Figure~\ref{fig:benchmark_spindles_consensus}. Indeed, at constant testing level, the F1 score at IoU = 0.3 decreases when the training consensus level increases. This might be due to the fact that fewer training and validation samples are available.
%
%
Note that, at constant training consensus level, the F1 score at IoU = 0.3 decreases when the testing consensus level increases on SSC. This might be explained by the fact that the events used for training are of poor quality compared to the events used for testing. This will be further commented in the discussion of this work.

Finally, on WSC, at constant training consensus level, the higher F1 score is always reached for a testing consensus level equal to the training one. This might be explained by the fact that the quality of annotated events for training and testing are similar.
%
%

\begin{figure*}[ht!]
\centering
\includegraphics[width=0.73\linewidth]{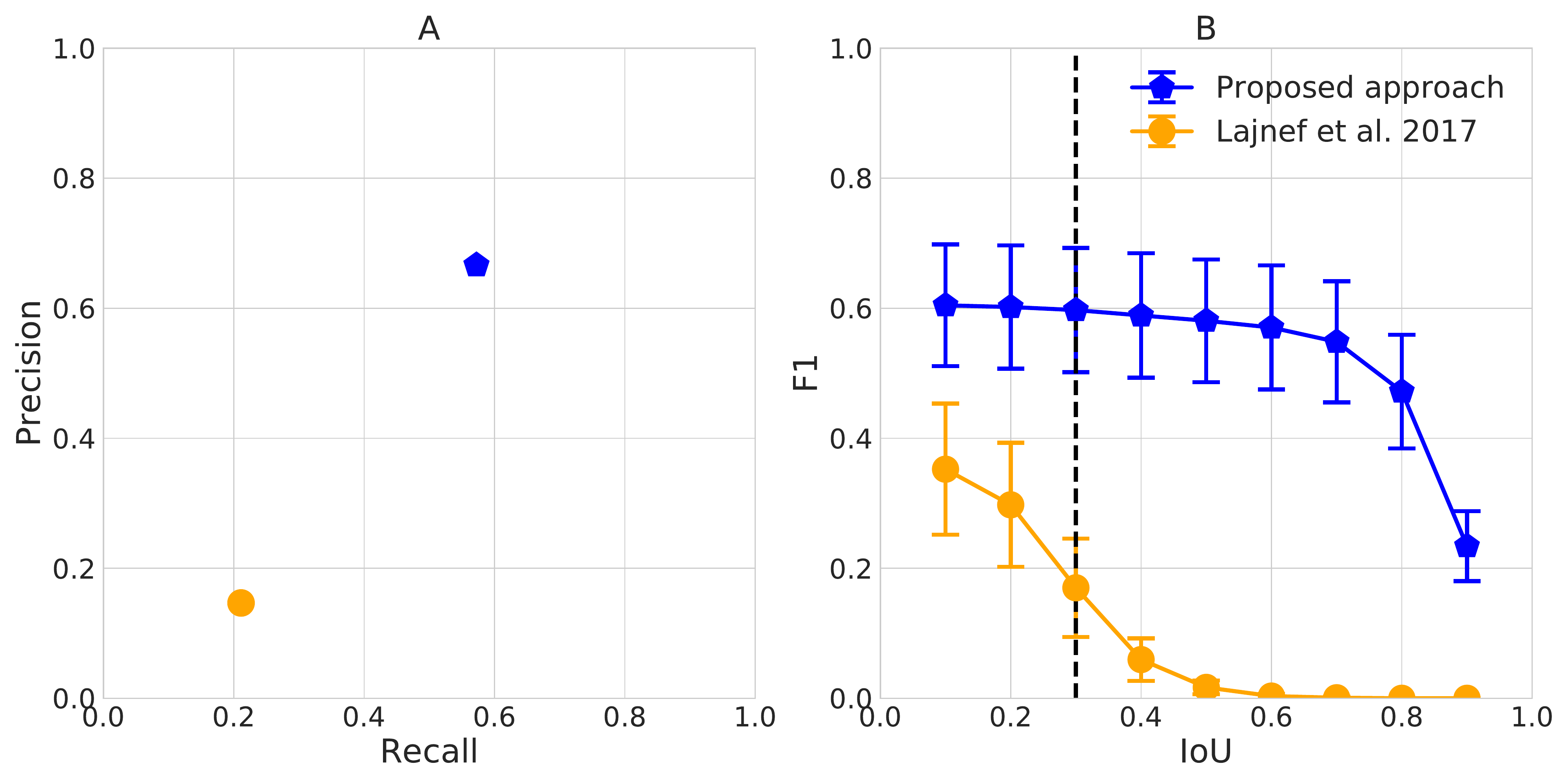}
\caption{K-complexes detection: the proposed approach outperforms the alternative approach~\cite{Lajnef2017}. A: Precision / Recall at IoU = 0.3. B: F1 score as a function of IoU.}
\label{fig:benchmark_kcomplexes}
\end{figure*}

\begin{figure*}[ht!]
\centering
\includegraphics[width=0.9\linewidth]{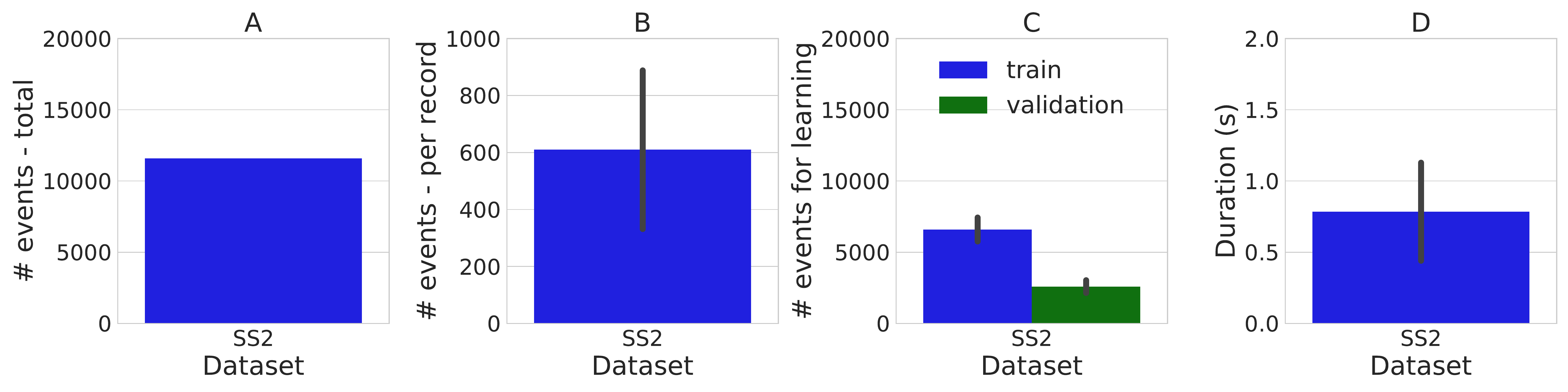}
\caption{Statistics about K-complexes on MASS SS2.  A: total number of events. B: averaged number of events per PSG recording. C: averaged numbers of events used for training and validation. D: averaged duration of events.}
\label{fig:statistics_kcomplexes}
\end{figure*}

\paragraph*{K-complexes detection}
In this experiment, we perform a general benchmark on MASS SS2 dataset, on K-complexes detection and demonstrate that the proposed approach outperforms current state-of-the-art methods~\emph{Lajnef et al. 2017}~\cite{Lajnef2017}. We report the obtained performances in Figure~\ref{fig:benchmark_kcomplexes}. We furthermore report some statistics about the K-complexes annotations over SS2 in Figure~\ref{fig:statistics_kcomplexes}.

The proposed approach outperforms the approach from \emph{Lajnef et al. 2017}~\cite{Lajnef2017} in terms of Precision / Recall at IoU = 0.3, and in terms of F1 score for any IoU. Furthermore, the proposed approach seems to predict start and end times quite precisely. Indeed, the F1 score as a function of IoU is rather stable for $\mathrm{IoU} \in [0.1, 0.7]$.
%
%
Note however, that \emph{Lajnef et al. 2017} is not able to predict accurately the start and end times of events and can only detect the negative peaks of K-complexes. Indeed, the start and end times are predicted empirically as occurring $0.1$\,s before and $1.3$\,s after the negative peaks of the K-complexes following the authors' pipeline~\cite{Lajnef2017}. This penalizes negatively this approach as it assumes that the duration of predicted events is always of $1.4$\,s whereas the duration of true events is around $0.8$\,s on average (see Figure~\ref{fig:statistics_kcomplexes}). This may explain the drop of F1 score for $\mathrm{IoU} \geq 0.3$.

\begin{figure*}[ht!]
\centering
\includegraphics[width=0.73\linewidth]{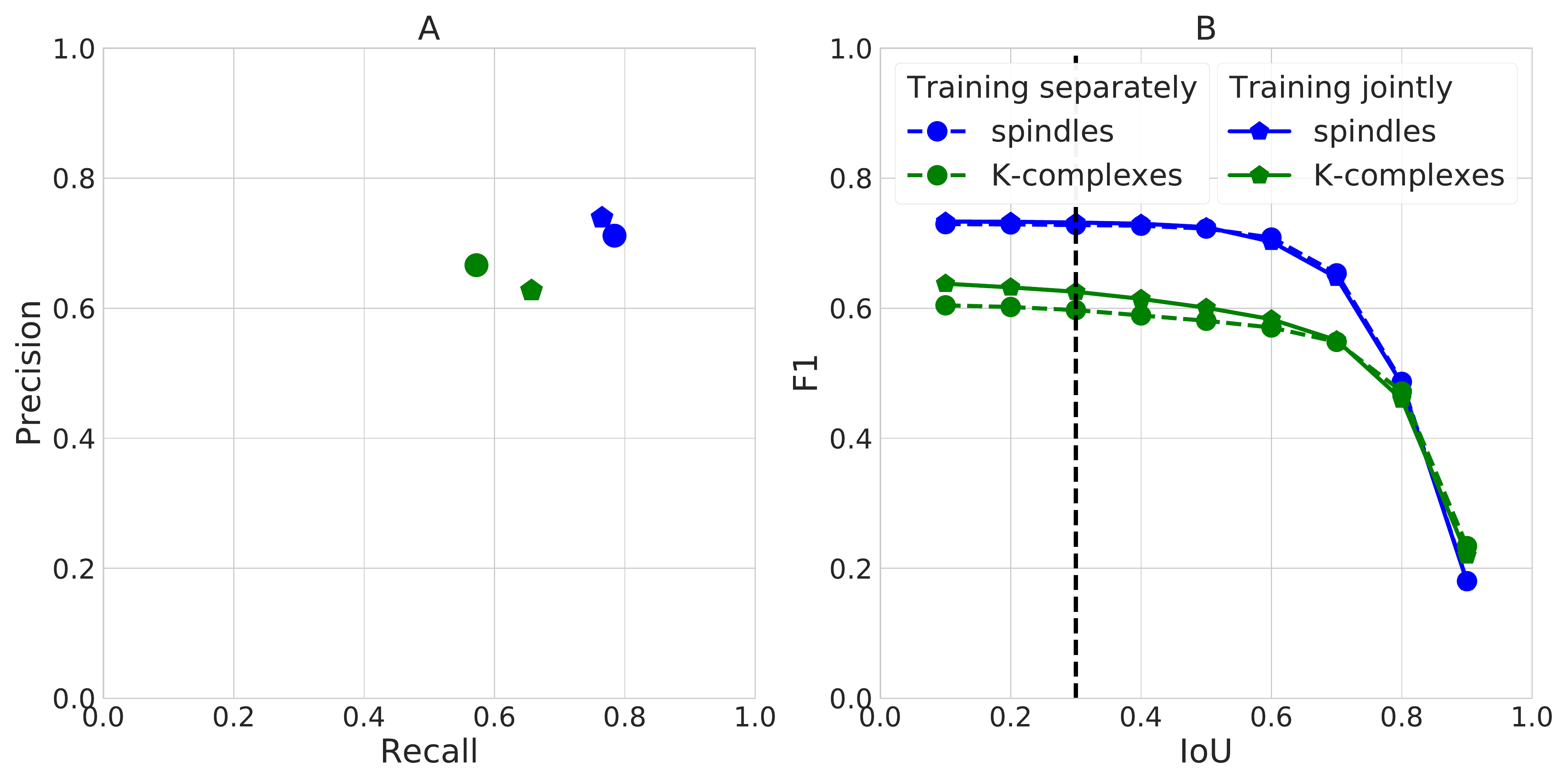}
\caption{Joint spindles and K-complexes detection: learning to detect both events jointly or separately leads to the same performances.}
\label{fig:benchmark_both_events}
\end{figure*}

\paragraph*{Joint spindles and K-complexes detection}
In this experiment, we perform a general benchmark on joint K-complexes and spindles detection and we demonstrate that detecting both events at the same time leads to detection performances similar to the ones obtained when performing each detection task separately. We report the results in Figure~\ref{fig:benchmark_both_events}.

Learning to detect both events jointly or separately leads to similar performances in terms of precision and recall at IoU = 0.3 and F1 score as a function of IoU.

\begin{figure*}[ht!]
\centering
\includegraphics[width=0.43\linewidth]{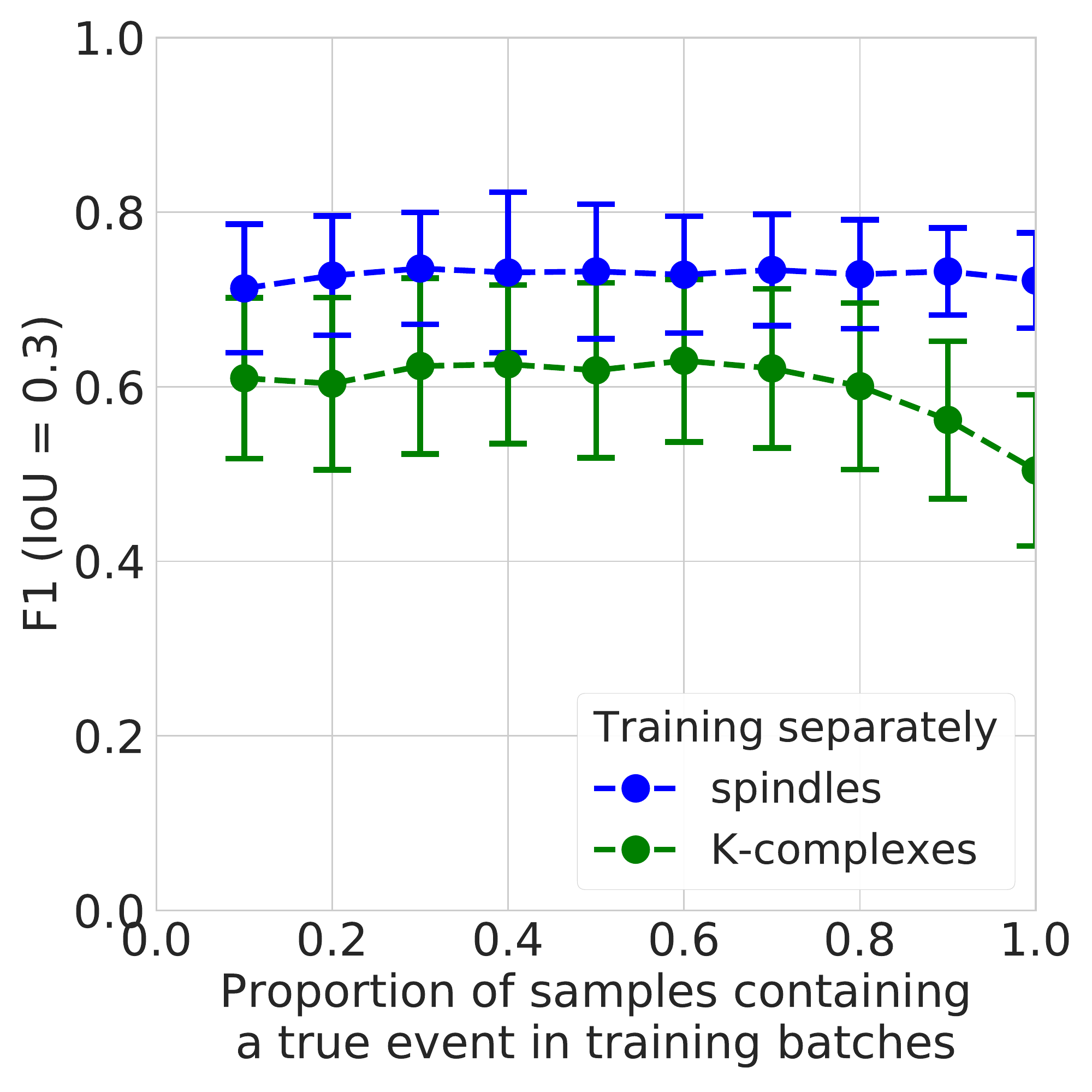}
%
%
\caption{Influence of batch sampling on detection performances: using batches only composed of samples containing a true event leads to lower performances than using batches composed of a mix of samples with a true events and "empty samples".}
\label{fig:sampling}
\end{figure*}

\paragraph*{Sampling}
In this experiment, we investigate the influence of the batch sampling strategy. We vary the proportion of samples in a training batch which contains a true event from $0.1$ to $1.$ and we quantify its influence on the F1 score at IoU = 0.3. In other words, setting the proportion to $0.5$ means that half of the samples present in the batch contains a true event. We report the obtained performances on SS2 dataset in Figure~\ref{fig:sampling}.

One can observe that the proportion of samples in a training batch, which contain a true event does not influence significantly the proposed approach's performance on spindles detection.

However, for K-complexes detection, a higher proportion of samples containing a true event in a batch leads to a significantly lower detection performance. This might be due to the following facts. First, K-complexes are events with a low frequency content arising in N2 sleep stage~\cite{iber2007} and similar low frequency content events also occur in N3 sleep stage under the form of grouped slow oscillations. Second, with high proportions of samples containing a true event the proposed approach is likely to never learn to differentiate a K-complex from a slow oscillation. The analysis of predictions with respect to sleep stages (not shown) agrees with this explanation. Indeed, a higher proportion (in training batches) of samples with a true event leads to a higher number of predicted K-complexes in period of N3 sleep. However no K-complex has been annotated over these periods of N3. To prevent such an issue, a $0.5$ proportion of samples containing a true event appears as a good compromise.

\begin{figure*}[ht!]
\centering
\includegraphics[width=0.98\linewidth]{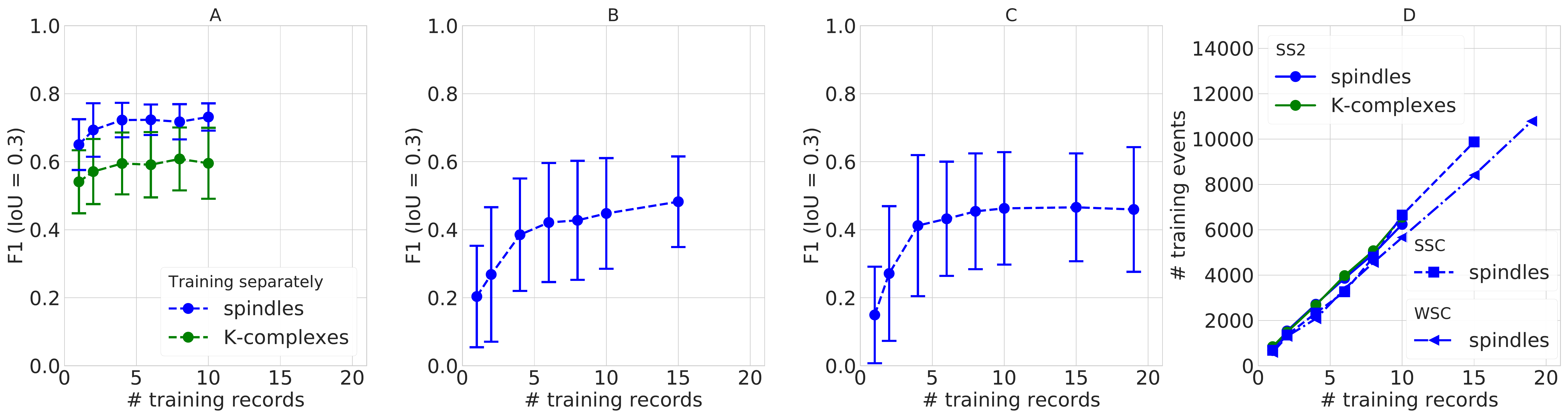}
\caption{Learning curves on SS2, SSC and WSC: more training PSG recordings leads to higher detection performances especially on SSC and WSC. A: SS2 - training separately. B: SSC. C: WSC. D: numbers of training events as a function of numbers of training PSGs for each dataset.}
\label{fig:learning_curves}
\end{figure*}

The same observations apply for the proposed approach when it is trained to detect jointly spindles and K-complexes (not shown).
%
%
\paragraph*{Learning curves: Do we have enough data?}
In this experiment, we investigate the influence of the quantity of labeled events used for training on the performances of the proposed approach on SS2, SSC and WSC. We report the F1 scores at IoU = 0.3 when the number of training PSG recordings varies from 1 to 20 (SS2: 1, 2, 4, 6, 8, 10, SSC: 1, 2, 4, 6, 8, 10, 15, WSC: 1, 2, 4, 6, 8, 10, 15, 20) in Figure~\ref{fig:learning_curves}. While computing these so-called \emph{learning curves}, the number of validation PSGs was kept fixed for each dataset.

%
%
The learning curves on SS2 demonstrate that learning from few PSG recordings leads to quite good performances. Indeed, the proposed approach exhibits high F1 at IoU = 0.3, even when trained on events from a single PSG. On the other hand, the learning curves on SSC and WSC lead to slightly different observations: learning from just a couple of PSGs is not sufficient to get good performances. Indeed, the performance reached when using the maximum number of training PSGs available is much higher than the performance obtained when using a few training PSGs. This experiment first indicates that the spindle detection task is more complex on SSC and WSC than on SS2 which agrees with observations previously made. Second, it suggests that we might lack of labeled data on SSC and WSC to obtain detection performances similar to the ones obtained on SS2.

\paragraph*{Arousal detection}
In this section, we apply and demonstrate that the proposed approach can be used to detect other types of events, here arousals. We investigate the following technical questions: (1) the influence of the duration of default events (2) the influence of the quantity of training data and (3) the influence of the number of channels considered.

\begin{figure*}[ht!]
\centering
\includegraphics[width=0.73\linewidth]{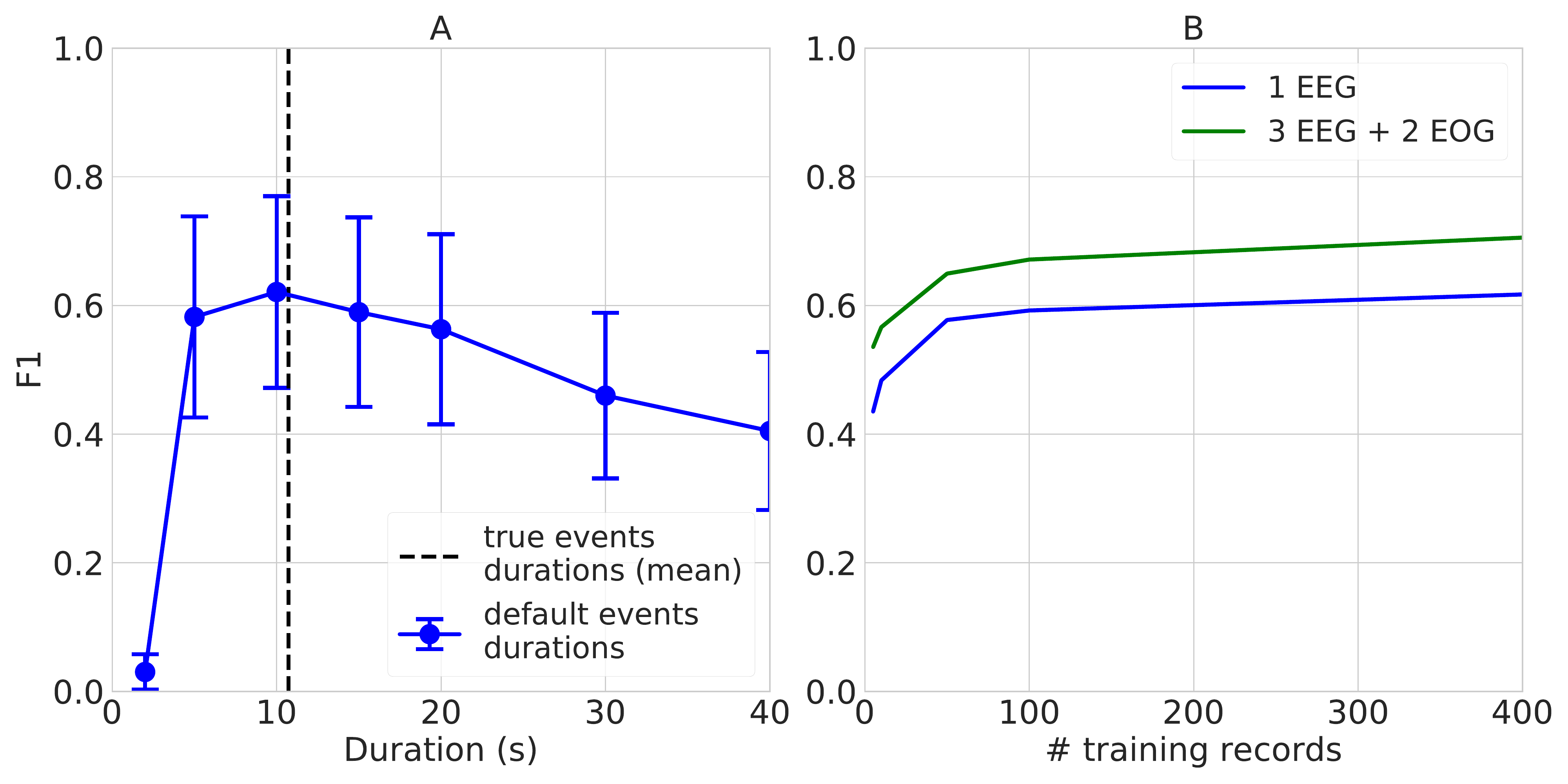}
\caption{Application of the proposed approach on arousals detection: the parameterization of default events appears as important as the quantity of training data or the quantity of spatial information. A: F1 score at IoU = 0.3 as a function of default event duration. B: F1 score at IoU = 0.3 as a function of the quantity of training PSG recordings and the number of channels}
\label{fig:duration_default_events}
\end{figure*}

\begin{figure*}[ht!]
\centering
\includegraphics[width=0.9\linewidth]{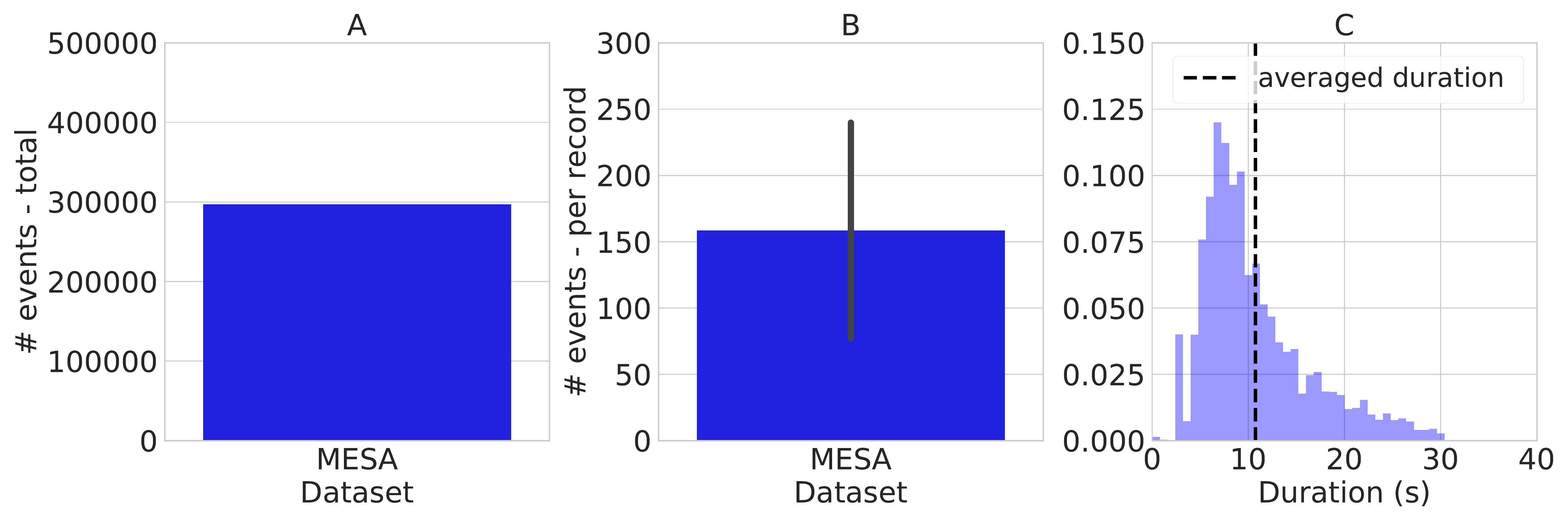}
\caption{Statistics about arousals in the $1000$ PSG recordings from MESA used for this detection task. A: total number of annotated events in the 1000 PSGs. B: averaged number of events per record. C: distribution of events durations}
\label{fig:stats_arousal}
\end{figure*}

To do so, we first train the proposed approach while varying the duration $t_i^d$ of default events from $2$\,s up to $40$\,s and keeping an overlapping of $50$\% between two consecutive default events: default events are generated every $t_i^d / 2$\,s ensuring that only $2$ default events overlap. We report the obtained results in Figure~\ref{fig:duration_default_events}~-~A. We furthermore investigate the influence of the quantity of training data available modulated by the quantity of spatial information that is processed, Figure~\ref{fig:duration_default_events}~-~B. We additionally report some statistics about the events annotated over MESA in Figure~\ref{fig:stats_arousal}. 

Three observations can be made from the reported results. First, scaling the duration of default events onto the averaged duration of true events leads to the best performances, see Fig~\ref{fig:duration_default_events}~-~A. This strategy shall be considered as the preliminary step when investigating the detection of new types of events with the proposed approach. Second, Figure~\ref{fig:duration_default_events}~-~B demonstrates that the more data the better, yet for this task and with the chosen network architecture, the metric increases at a limited pace after 100 training PSG recordings. Finally, using more channels leads to a significant boost in detection performances and may also be used to balance a potential lack of training PSGs as shown in Figure~\ref{fig:duration_default_events}. This positive impact of the number of channels was also reported in~\cite{Chambon2018}.


\section{Discussion}
This work reports on the successful use of a novel approach to simultaneously detect multiple EEG microarchitectural features in EEG signals.  It is based on object recognition approaches developped for computer vision and is inspired by SSD~\cite{Liu2016} and YOLO approaches~\cite{Redmon2015}. The model builds by back-propagation a feature representation of the data relevant for the task to perform and learns how to predict centers, durations and classes of events of interest. The approach is versatile enough to detect any type of events provided some labeled samples are available. A major advantage of the proposed approach compared to state-of-the-art baselines is that it can detect multiple events of multiple time scales simultaneously. It could therefore easily be extended to detect other types of events in other modalities or concurrent signals (for example sleep disordered breathing events on breathing channels or Periodic Leg Movements during sleep on leg EMGs), provided labeled events are available. This work also raises interesting technical issues that are essential to a successful application of this algorithm, such as the importance of including an heterogeneous set of clinical PSGs of different origins, having sufficient data available for training, the importance of correct labeling and of defining optimal duration for default events.

\paragraph*{Impact of clinical population and labeling errors}

The importance of datasets and of proper labeling is best illustated by the significant gap in performances observed between SS2 and SSC / WSC datasets in detecting spindles using all considered methods (see also Figure~\ref{fig:learning_losses} in Appendix). One factor that may explain the differences in performance is related to the saliency of the frequency content inside the $11 - 16$\,Hz band. Indeed, power spectrum in this band is much higher for SS2 than SSC / WSC, see Figure~\ref{fig:statistics_spindles}~-~E. The frequency difference might be due to population differences between SS2 and SSC / WSC datasets. Indeed, subjects in SSC / WSC are older, and it is known that age affects spindle frequency, duration, density and amplitude~\cite{Purcell2017}. These age dependent changes may make spindle detection harder for experts and for algorithms.  It also illustrates the need to test performance for any new method on multiple datasets.  

Another likely factor that affects performance across datasets is quality of experts annotations. Indeed, some scorers using the SSC and WSC data were unable to precisely mark start and end times of events (up to a 0.5\,s precision) due to PSG viewer limitations. Annotations can depend on guidelines given to scorers, their interpretations, and other technical issues such as the viewer used~\cite{Warby2014}.  Taking the union of scorers annotations allowed us to partially cope with this problem but not entirely. Indeed, there is no strong guarantees that a consensus strategy will improve annotations quality. Besides, a higher consensus level leads to fewer annotated events making it harder to train using the proposed approach (see Figure~\ref{fig:statistics_spindles_consensus}).  One possible strategy to improve on this issue could be to use a larger number $N_s$ of scorers, as previously performed~\cite{Warby2014}, and selecting only the $K \in \N$ best scorers depending on overall performances compared to a consensus of $N_s - 1$ scorers.
\paragraph*{Quantity of available data for training}
More data can mean different things. It can mean more PSG recordings or more channels, for example with more modalities than just EEG. To address the question of the impact of the number of PSGs, we computed learning curves which present the prediction performance as a function of the number of training samples (Figure~\ref{fig:learning_curves}). Surprisingly, this analysis shows that training with the proposed approach to detect spindles or K-complexes on just $1$ PSG already leads to decent performances on SS2. This result is likely explained by the sampling strategy used in the training process. During training, windows are sampled at random time in the PSG recording. This process is thus unlikely to input twice the exact same signal to the network while training, hence limiting the risk of over-fitting. Figures~\ref{fig:learning_curves}~-~B~/~C also demonstrate that the proposed approach benefits significantly more from the availability of additional PSGs for SSC and WSC. This suggests that the learning task is more difficult on these two datasets. This can be explained, as already discussed above, by the clinical nature of the population and annotations present in these two datasets.  A similar comment applies to the detection of arousals, as F1 score at IoU = 0.3 increases as a function of the number of training PSG recordings, and reaches a smaller increase pace from $100$ training PSGs in Figure~\ref{fig:duration_default_events}~-~B.

Regarding the quantity of spatial information, using multiple EEG and EOG channels for arousals detection delivers a significant gain of performance. This demonstrates that increasing the number of channels can compensate for a potential lack of training PSG recordings (see Figure~\ref{fig:duration_default_events}~-~B). Investigating the joint processing of multiple modalities, such as electro-myography, breathing, or pulse-oxymetry signals remains to be done for event detection, although one can expect an increase in performance~\cite{Chambon2018,Olesen2018,Andreotti2018}. In particular, such a perspective might prove successful to detect specific types of events occurring through multiple modalities like sleep-apnea.

\paragraph*{Impact of default event duration and joint detection of multiple events with variable durations}
The parameterization of default events is a crucial step of the proposed approach. While the parameterization of default events for spindles and K-complexes detection was quite \emph{straightforward} because these events exhibit a 1\,s averaged duration, experiments on arousals detection show that default duration and overlap factor must be carefully selected. Indeed, bad parameterization compromises training, as some true events are never matched during training and are thus never used to train the network localization and prediction modules. This reduces the effective number of samples considered for training and makes recognition of such events more difficult. The selection of a good default duration is a crucial first step when investigating the use of the proposed approach for a new type of event. A good heuristic though is to set the duration of default events to the averaged duration of events in the training set. Of note, detecting jointly multiple events is possible with the proposed model as demonstrated by Figure~\ref{fig:benchmark_both_events}. However it shall be stressed that in this experiment, we explored the joint detection of spindles and K-complexes, features which exhibit about the same durations. To detect events of different durations, \emph{e.g.} spindles and arousals for example, one would need to parameterize default events of multiple scales which was not investigated in this work.

\paragraph*{The influence of the model architecture and of balanced sampling on performance}
The proposed approach exhibits a VGG-like architecture~\cite{Simonyan14}, where for each convolution block, the number of feature maps is multiplied by $2$ while the temporal resolution is divided by $2$. Such a choice enables to extract relevant features for the considered tasks, although more complex architectures remain to be investigated. The proposed approach exhibits much higher performances than our earlier work~\cite{Chambon2018a}. This stems from two major changes.

The first major change is related to the network architecture. The implementation of prediction modules in this work differs from the one embraced in our previous work~\cite{Chambon2018a} as we have opted for predicting locations and classes of default events from the whole feature maps returned by $\phi_T$. This choice had three positive impacts: (1) it boosted the detection performances of K-complexes and (2) it allowed for a simpler parameterization of defaults events, making them independent of the size of the feature maps returned by $\phi_T$, also allowing to freely choose numbers, durations and overlaps of defaults events to use. Finally, (3) it also allowed to predict a wide range of events durations: 1\,s spindles as well as 10\,s arousals with the same architecture - provided one took care of the parameterization of default events as mentioned above. Our previous approach was predicting potential events by processing $3$ consecutive time steps of the feature maps returned by $\phi_T$ which may not give access to a sufficiently large temporal context for the prediction of long events like arousals.

The second change is the use of a balanced sampling strategy during training. The current approach, training batches are composed of 50\% of samples containing no event of interest. This led to a significant increase in performance for the detection of K-complexes (see Figure~\ref{fig:sampling}). It also contributed to a reduction of false positives predicted in N3 sleep, which resulted in a boost of F1 score.




\paragraph*{Comparison with YOLO and SSD and Perspectives}
Our algorithm is inspired by both the SSD~\cite{Liu2016} and YOLO approaches~\cite{Redmon2015}. Similarities between our work and these approaches include the fact that we employ an SSD loss function which is a combination of a smooth L1 loss and a classification loss. Furthermore, the same matching strategy as SSD is used during training. Finally, the prediction module $\psi$ makes predictions from the whole feature map returned by the feature extractor $\phi_T(\phi_C)$. This is similar to YOLO. To that end, the proposed approach relies on a grid of default events independent of the temporal size of this latter features maps. This allows for more flexibility in the definition of default event.  

Our proposed approach howewer also differs from SSD and YOLO because it predicts on large objects, such as a 10\,h PSG containing mostly no events. The approach was thus further developed to processes chunks of signals and trained both on samples containing true events and on samples containing no event of interest (see balanced sampling strategy discussed above). 

The proposed approach could be extended into several directions. First, the model actually predicts from a single level of feature maps. Exploiting several levels of feature maps containing multiple temporal resolutions and associated with several scales of default events as performed by SSD~\cite{Liu2016}, or more recent approaches such as Feature Pyramid Networks~\cite{lin2016fpn} or RetinaNet~\cite{Lin2017} could be considered. This would likely boost detection performances by leveraging detection of events which exhibit a wide range of durations (from 1\,s to possibly several minutes events), this being espacially useful in the context of full PSG automatic scoring of both macro and microarchitectural events. Second, exploiting more complex prediction modules, combining several convolution layers and non-linearities, could enhance detection performances ~\cite{Lin2017,lin2016fpn}. Third, minimizing a different loss function like Focal Loss~\cite{Lin2017} should also be considered as a direction of investigation.  Finally, events of interests such as spindles might exhibit specific temporal dynamic patterns, with the likelihood of a spindle occurring in a sample being related to the occurrence of spindles in a previous sample. Integrating a temporal context using a recurrent neural network as performed for EEG processing~\cite{Bashivan2016} or for sleep stage classification~\cite{Dong2016,Supratak2017,Phan2018, Stephansen2017,Phan2019b} might enhance detection performance for some events.  In all cases, however, the proposed approach has the considerable advantage of simultaneous multi-event detection, a crucial feature that should allow to build more easily additional event detection methods on the same architecture.

\section*{Acknowledgments}
This work was supported in part by the french Association Nationale de la Recherche et de la Technologie (ANRT) under Grant 2015 / 1005.

\section*{References}
\bibliographystyle{elsarticle-num}
\bibliography{biblio}


\section{Appendix}

\subsection{Benchmark spindles - learning losses}
\begin{figure*}[ht!]
\centering
\includegraphics[width=0.78\linewidth]{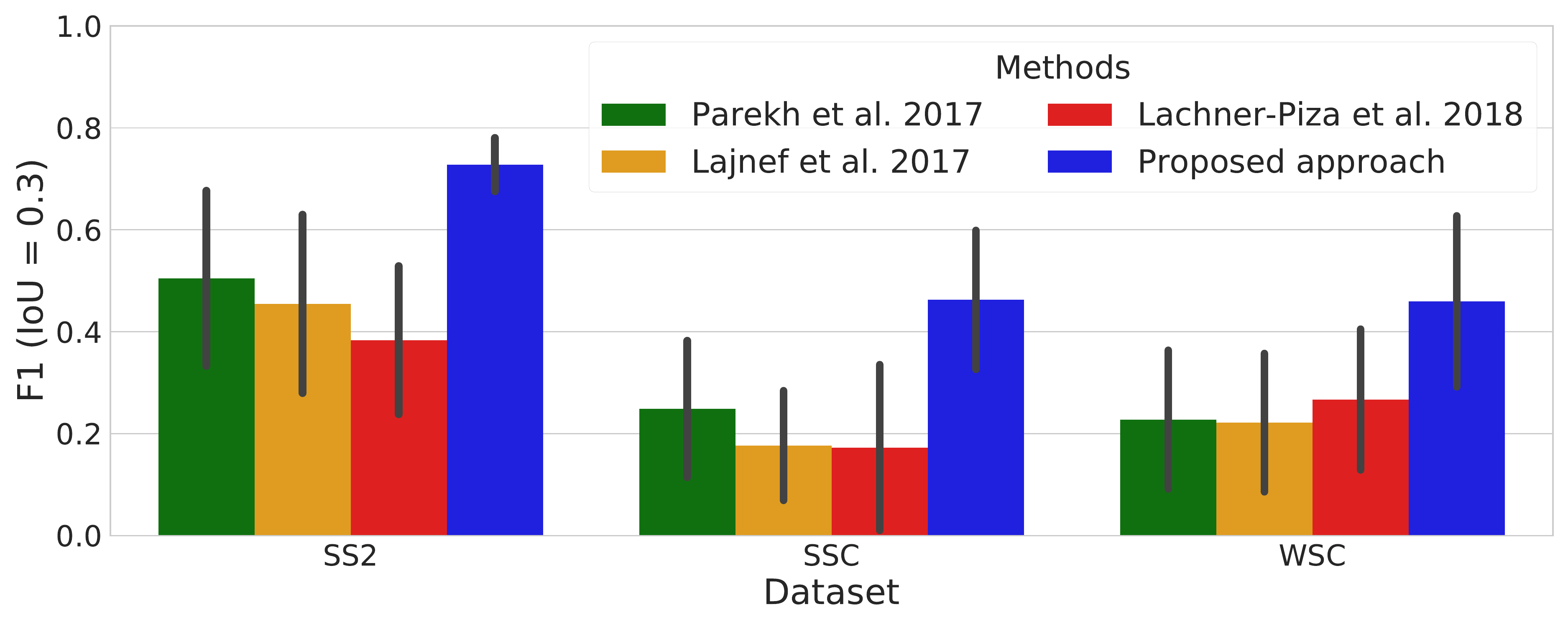}
\caption{Averaged F1 scores at IoU = 0.3 and standard deviations: on SSC and WSC the proposed approach exhibits lower detection performances and larger standard deviations}
\label{fig:benchmark_spindles_sup}
\end{figure*}

\begin{figure*}[ht!]
\centering
\includegraphics[width=0.88\linewidth]{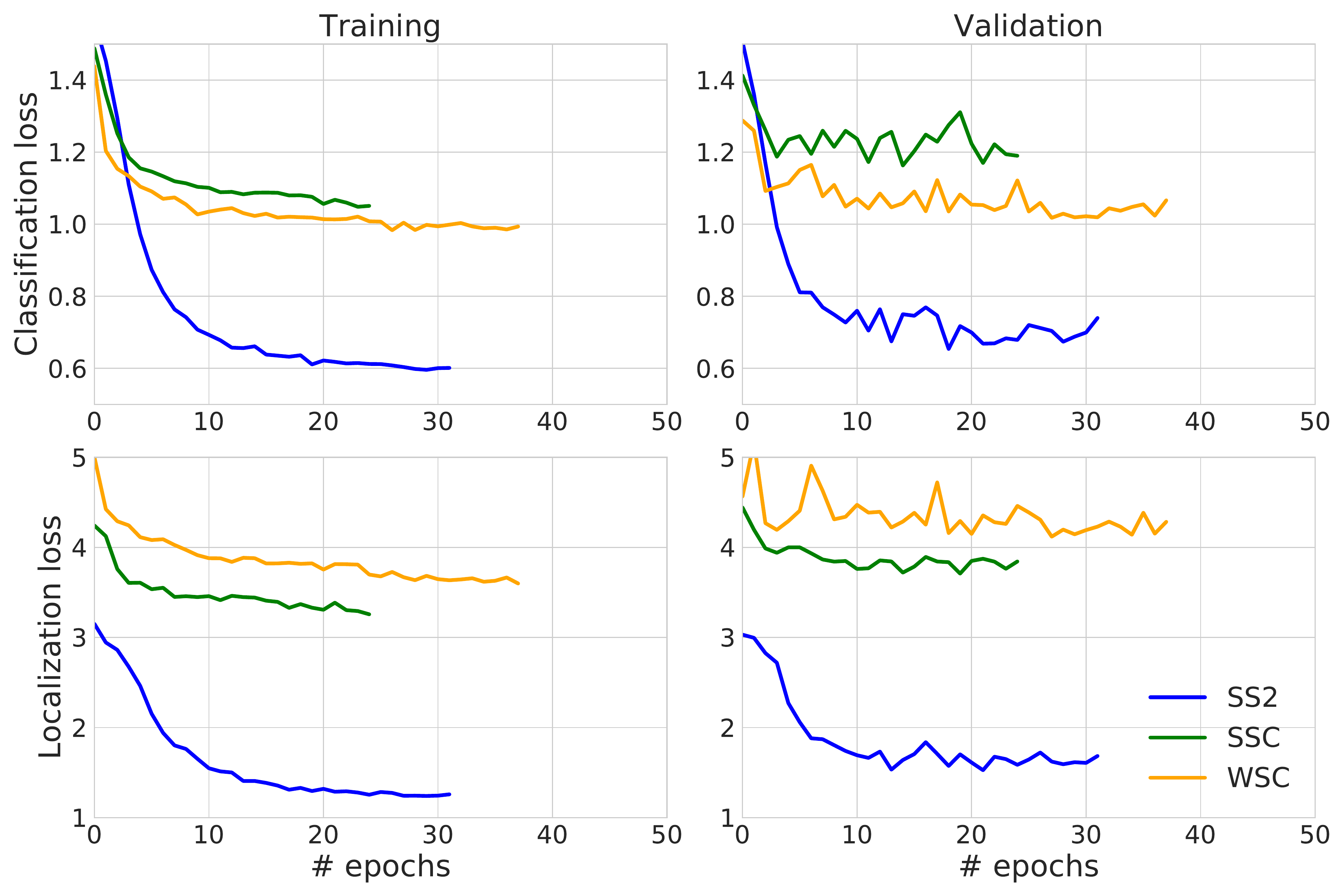}
\caption{Classification and localization losses for spindles detection, on training and validation sets, over the first split of cross-validation for each dataset: the losses agree with the assumption that the spindles detection task is easier on SS2 than on SSC / WSC}
\label{fig:learning_losses}
\end{figure*}

In this paragraph, we investigate further the performances obtained by the proposed approach on the spindles detection task, over the 3 considered dataset: SS2, SSC and WSC. We demonstrate that the spindle detection task is easier for the proposed approach on SS2 compared to SSC and WSC. We report in Figure~\ref{fig:benchmark_spindles_sup} the F1 scores at IoU = 0.3 and the standard deviations in scores obtained by each method. In Figure~\ref{fig:learning_losses}, we present the classification and the localization losses (on the training and the validation sets) during the learning for the first split over each dataset.

The detection scores reached by the proposed approach exhibit large differences between the 3 considered datasets: the averaged scores are higher on SS2 than on SSC and WSC, and the standard deviations are smaller. The fact that the state-of-the-art baselines suffer also some decreases in their detection performances suggests that the spindles detection task is more difficult on SSC and WSC than on SS2.

The classification and localization losses during training agree with this assumption. Indeed, the classification and localization losses exhibited by the proposed approach on SS2 reach much lower asymptotic values than on SSC or WSC. This indicates that the events are much easier to identify and localize on SS2 than on SSC or WSC.

\end{document}